\documentclass[prl,twocolumn]{revtex4}
\usepackage{amssymb}
\usepackage{amsmath}
\usepackage{bm}

\newcommand{\xx}{\mathbf x}
\newcommand{\pp}{\mathbf p}
\newcommand{\kk}{\mathbf k}
\newcommand{\BB}{\mathbf B}
\newcommand{\bb}{\mathbf b}
\newcommand{\om}{\bm \omega}

\begin{document}

%\twocolumn[{%
%\centering
% \Large \textbf{Chiral Magnetic-Vortical Wave} \\[1.5em]
% \large D.~Frenklakh 
%
%\small Institute of Theoretical and Experimental Physics, B. Cheryomushkinskaya 25, Moscow 117218, Russia \\
%\small Moscow Institute of Physics and Technology, Dolgoprudny 141700, Russia\\[0.5cm]
%
%\small ~~~~~~~~~~~~~~~We study collective excitations in rotating chiral media in presence of magnetic~~~~~~~~~~~~~~~ 
%
%field both in hydrodynamic framework and in kinetic theory. We find that the 
%
%velocity of the mixed Chiral Magnetic-Vortical Wave is a vector sum of 
%
%velocities of pure Magnetic and Vortical waves which do not exist separately 
%
%under these conditions. We also use relaxation time approximation to prove 
%
%that this wave itself is a non-dissipative phenomenon. \\[0.5cm]
%}]

\author{D.~Frenklakh}

\affiliation{B. Cheremushkinskaya, 25, Institute of Theoretical and Experimental Physics, Moscow 117218, Russia \\ Institutskii per, 9, Moscow Institute of Physics and Technology, Dolgoprudny 141700, Russia}

\title{Chiral magnetic-vortical wave}

\begin{abstract}
We study collective excitations in rotating chiral media in presence of magnetic field both in hydrodynamic framework and in kinetic theory. We find that the velocity of the mixed Chiral Magnetic-Vortical Wave is a vector sum of velocities of pure Magnetic and Vortical waves which do not exist separately under simultaneous presence of rotation and magnetic field. We also use relaxation time approximation to prove that this wave is a non-dissipative phenomenon in linear order in $B$ and $\omega$.
\end{abstract}

\maketitle

\section{Introduction}
Transportation effects driven by anomalies in chiral systems have attracted large interest recently. They can by induced by external magnetic field like in the Chiral Magnetic Effect(CME) , where electric current is generated along the magnetic field due to the presence  of axial chemical potential(\cite{ref:CME}). In magnetic field there also exists the Chiral Separation Effect (CSE), where conversely axial current along the magnetic field is generated in presence of  vector chemical potential (\cite{ref:CSE1, ref:CSE2}). Another example is the Chiral Vortical Effect(CVE) (\cite{ref:CVE1,ref:CVE2,ref:CVE3,ref:CVE4}) occurring in rotating chiral systems. 

These anomalous transport effects couple vector and axial charge densities and currents which leads to the existence of gapless excitations such as the Chiral Magnetic Wave (CMW)(\cite{ref:CMW}) and the Chiral Vortical Wave (CVW)(\cite{ref:CVW}). One may wonder what excitation could occur in the rotating chiral system placed in the external magnetic field. In hydrodynamic framework it was recently investigated in (\cite{ref:CHW}). In this paper we briefly discuss both linear and non-linear case in the hydrodynamic approach and then proceed to the analysis in chiral kinetic theory. 
% Hydrodynamic approach

\section{Hydrodynamic approach}
In this section we, in general, follow (\cite{ref:CHW}). We  consider the rotating system of both right and left Weyl fermions placed in constant homogeneous external magnetic field $\BB$. We assume the angular velocity of the system $\om$ to be constant.  The expressions for chiral currents are easily obtained from CME, CSE and CVE:
\begin{equation}
\mathbf j_{R/L} = \pm\frac{1}{4\pi^2}\mu_{R/L}\BB \pm \left(\frac{1}{4\pi^2}\mu_{R/L}^2 + \frac{1}{12}T^2\right)\om ,
\end{equation}
where $\mathbf j_{R/L} = \frac{1}{2}(\mathbf j_V\pm\mathbf j_A)$~($\mathbf j_V$ is vector current and $\mathbf j_A$ is axial current) and $\mu_{R/L} = \mu\pm\mu_5$. The corresponding charge densities are $n_{R/L} = \frac{1}{2}(j_V^0\pm j_A^0) .$

The continuity equations read as
\begin{equation}
\partial_t n_{R/L} + \bm\nabla\cdot\mathbf j_{R/L} = 0 .
\end{equation}
We consider small fluctuations of densities $\delta n_{R/L}$ on top of a uniform equilibrium background assuming the temperature to be constant. Les us choose the axes so that $\BB = (B,0,0)$ and $\om = (\omega_1,\omega_2,0)$. Then
\begin{multline}\label{hydro}
\partial_t\delta n_{R/L} \pm\frac{1}{4\pi^2}(B\partial_x\delta\mu_{R/L}+\omega_1\partial_x\delta(\mu_{R/L})^2 \\
 + \omega_2\partial_y\delta(\mu_{R/L})^2) = 0 .
\end{multline}
 We introduce susceptibilities for the corresponding densities $\chi_{R/L} = \frac{\partial n_{R/L}}{\partial \mu_{RL}}$. Let us concentrate on the right-handed particles below. If in equilibrium $\mu_R = \mu_0$, then up to the first order $\delta\mu_R^2 = \dfrac{2\mu_0}{\chi_R}\delta n$.  So (\ref{hydro}) transforms into
\begin{equation}
\partial_t\delta n + \frac{1}{4\pi^2\chi_R}((B+2\omega_1\mu_0)\partial_x\delta n + 2\omega_2\mu_0\partial_y\delta n) = 0 .
\end{equation}
The corresponding velocity of the wave described by this linear equation is 
\begin{equation}
v_R = \frac{|\BB + 2\mu_0\om|}{4\pi^2\chi_R} .
\end{equation}

This shows that velocities of CMW and CVW simply sum up as vectors. Note that if $\BB + 2\mu_R\om = 0$ the velocity is zero. Let us investigate this case more deeply. We need to take into account the non-linear terms that we have dropped before. Note that
\begin{equation}
\delta{\mu_R^2} = 2\mu_0\delta\mu_R + (\delta\mu_R)^2 .
\end{equation}
Substituting into~(\ref{hydro}) ($\omega_2=0$ and $B+2\mu_0\omega = 0$ now) we have
\begin{eqnarray}
\partial_t\delta n_R + \frac{1}{4\pi^2}(B+2\mu_0\omega)\partial_x\delta\mu_R +\frac{\omega\partial_x(\delta\mu_R)^2}{4\pi^2} \nonumber \\
=\partial_t\delta n_R + \frac{\omega\partial_x(\delta n_R)^2}{4\pi^2\chi_R^2} = 0 .
\end{eqnarray}
This is Hopf equation and notably magnetic field has disappeared from the expression, so in this case the features of the solution are defined only by vorticity. The implicit solution of this equation is 
\begin{equation}
\delta n(x,t) = F\left(x-\frac{\omega t\delta n}{2\pi^2\chi_R^2}\right) ,
\end{equation}
where $F(x)=\delta n(x,t=0)$ is initial density distribution.

%Kinetic approach
\section{Kinetic approach}
Kinetic theory recently has been used to study transportation processes such as CME and CVE as well as the corresponding gapless excitations - CMW and CVW (\cite{ref:Kinetic1,ref:CVW}). 
In this section we study the single right-handed Weyl fermion field. While quantization of such a field right particles and left antiparticles emerge, they will be denoted by indices + and - respectively. We consider the temperature to be high (compared to $\sqrt B$, $\omega$ and background chemical potential $\mu$ which is necessary for treating fermions classically with the only exception for Berry connection, see \cite{ref:Kinetic2}) and constant. Taking similar approach as in (\cite{ref:CVW, ref:Kinetic1}) we start from the kinetic equation
\begin{equation}\label{kinetic}
\frac{\partial f_{\pm}}{\partial t} + \dot\xx\cdot\frac{\partial f_{\pm}}{\partial \xx}+\dot\pp\cdot\frac{\partial f{\pm}}{\partial \pp} = C_{\pm} [f_+,f_-] ,
\end{equation}
where $f_{\pm}(t,\xx,\pp)$ are distribution functions and $C_{\pm}[f_+,f_-]$ are collision integrals. 
The equations of motion for particles $(+)$ in their local rest frame are
\begin{equation}
\dot\xx = \hat\pp + \dot\pp\times\bb , ~\dot\pp = \dot\xx\times(\BB+2p~\om) ,
\end{equation}
similarly to (\cite{ref:Kinetic2}) with additional Coriolis force term. Here $p = |\pp|$, $\hat\pp = \dfrac{\pp}{p}$ and $\bb = \dfrac{\hat\pp}{2p^2}$ is the curvature of Berry's connection in momentum space. From these equations(and similar equations for antiparticles) we obtain
\begin{eqnarray}
\sqrt{G_{\pm}}\dot\xx = \hat\pp + \BB'_{\pm}(\hat\pp\cdot\bb) ,\\
\sqrt{G_{\pm}}\dot\pp = \pm\hat\pp\times\BB'_{\pm} ,
\end{eqnarray}
where $\sqrt G_{\pm} = 1+\BB'_{\pm}\cdot\bb$ modifies phase space volume due to the interplay between effective magnetic field and Berry connection, $\BB'_{\pm} = \mathbf B \pm 2 p ~\om$ is the effective magnetic field. 

We now want to linearise the kinetic equation. Consider small fluctuations above the equilibrium Fermi-Dirac distribution $f_{0\pm}(p)$:
\begin{equation}
f_{\pm}=f_{0\pm}(p)-\partial_p f_{o\pm}(p)\delta f_{\pm}(t,\xx,\pp) ,
\end{equation}
and take the Fourier transformation of $\delta f_{\pm}(t,\xx,\pp)$ to be $h_{\pm}(\nu,\kk,\pp)$. In equilibrium the collision term is zero, so we can write it as
\begin{equation}
C_{\pm}[f_+,f_-] = -\partial_p f_{0\pm}I_{\pm}[h_+,h_-]+O(h^2) .
\end{equation}
The kinetic equation (\ref{kinetic}) now becomes 
\begin{equation}\label{reduced}
-i\nu h_{\pm} + \dot\xx(i\kk \pm \BB_{\pm}\times\frac{\partial}{\partial\pp})h_{\pm} = I_{\pm}[h_+,h_-] .
\end{equation}
Now we want to average this in the momentum space, so we define the brackets $\langle...\rangle_{\pm}$ as
\begin{equation}
\langle...\rangle_{\pm} = \int_{\pp}\sqrt{G_{\pm}}\partial_p f_{0\pm}(p)(...) ,
\end{equation}
 where $\int_{\pp} = \int\dfrac{d^3p}{(2\pi)^3}$. From the charge conservation constraint one obtains 
\begin{equation}
\int_{\pp}\sqrt{G_+}C_+[f_+,f_-]-\int_{\pp}\sqrt{G_-}C_-[f_+,f_-] = 0 ,
\end{equation}
for arbitrary $f_{\pm}$ which implies 
\begin{equation}
\int_{\pp}\sqrt{G_+}\partial_p f_{0+} I_+[h_{\pm}] - \int_{\pp}\sqrt{G_-}\partial_p f_{0-} I_-[h_{\pm}] = 0 ,
\end{equation}
for arbitrary $h_{\pm}$.
Also, the "Lorentz force" term vanishes after averaging and integrating by parts.
So, averaging the equations (\ref{reduced}) with the corresponding bracket and taking the difference we obtain
\begin{eqnarray}\label{empty}
\nu(\langle h_+\rangle_+ - \langle h_-\rangle_-) - \kk(\langle\dot\xx h_+\rangle_+ - \langle\dot\xx h_-\rangle_-) = 0 .
\end{eqnarray}
\subsection{Hydrodynamic regime}
We now want to study hydrodynamic regime, which similarly to \cite{ref:CVW} means that we consider $h_{\pm}$ to be independent on $p$ and $h_+ = -h_- = h$. This means $h_{\pm}$ can be taken out of averaging brackets and the equation becomes
\begin{equation}{\label{kinhydro}}
2\nu h(\langle1\rangle_+ + \langle1\rangle_-)-\kk(\langle\dot\xx\rangle_+ + \langle\dot\xx\rangle_-)h = 0.
\end{equation} 
For $\langle\dot\xx\rangle_+$ we have
\begin{multline}
\langle\dot\xx\rangle_+ = \int_{\pp}(\hat\pp + \BB'_+(\hat\pp\cdot\bb))\frac{\partial f_{0+}(p)}{\partial p} = \int_{\pp}\BB'_+(\frac{\partial f_{0+}(p)}{\partial \pp}\cdot\bb) \\ 
= -\int_{\pp}\BB'_+(f_{0+}\frac{\partial b_i}{\partial p_i}) -\int_{\pp}f_{0+}(b_i\frac{\partial\BB'_+}{\partial p_i}) \\ 
= -\int_{\pp}f_{0+}(p)2\pi\delta^{(3)}(\pp)\BB'_+(p) -\int_{\pp}f_{0+}(b_i(2\hat p_i\om)) \\ 
= -\frac{\BB}{8\pi^2} - \om\int_{\pp}\frac{f_{0+}}{p^2} = -\frac{\BB}{8\pi^2} - \frac{\om}{2\pi^2}\int_0^\infty f_{0+}dp .
\end{multline}
For Fermi-Dirac distribution $f_{0\pm} = \dfrac{1}{e^{\beta(p\mp\mu_0)}}$ we get
\begin{equation}
\int_0^\infty f_{0+}dp = \frac{1}{\beta}\log(1+e^{\beta\mu_0}) .
\end{equation}
Analogously
\begin{eqnarray}
\langle\dot\xx\rangle_- = -\frac{\BB}{8\pi^2} + \frac{\om}{2\pi^2}\int_0^\infty f_{0-}dp \\
\int_0^\infty f_{0-}dp = \frac{1}{\beta}\log(1+e^{-\beta\mu_0}) .
\end{eqnarray}
Finally 
\begin{multline}{\label{sum}}
\langle\dot\xx\rangle_+ + \langle\dot\xx\rangle_- = - \frac{\BB}{4\pi^2} - \frac{\om}{2\pi^2\beta}\log\left(\frac{1+e^{\beta\mu_0}}{1+e^{-\beta\mu_0}}\right)  \\ 
\approx -\frac{\BB}{4\pi^2} - \frac{\om\mu_0}{2\pi^2} .
\end{multline}
in our approximation $\beta\mu_0\ll1$. 
\begin{multline}{\label{susc}}
\langle1\rangle_+ + \langle1\rangle_- = \int_{\pp}\sqrt{G_+}\partial_pf_{0+} + \int_{\pp}\sqrt{G_-}\partial_pf_{0-} \\
 = - \frac{\partial n}{\partial\mu}|_{\mu_0} = - \chi ,
\end{multline}
where $\chi$ is charge susceptibility. So from (\ref{kinhydro},\ref{sum},\ref{susc}) we get the velocity of the Chiral Magnetic-Vortical Wave
\begin{equation}
\mathbf v = \frac{\BB + 2\mu_0\om}{4\pi^2\chi} .
\end{equation}

Notably this result coincides with the one obtained in \cite{ref:CHW} and reobtained in the previous section. 

\subsection{Relaxation time approximation}
We now want to study the effect beyond the hydrodynamic regime and use relaxation time approximation (RTA) for the collision term like in \cite{ref:Kinetic1}. The collision term in this approximation is
\begin{equation}
I_{\pm}[h_+,h_-] = -\frac{1}{\tau}(h_{\pm}\mp\bar h) ,
\end{equation}
where from the charge conservation
\begin{equation}
\bar h = \frac{1}{2}\langle h_+\rangle_+ - \langle h_-\rangle_-) .
\end{equation}
We consider $\BB$ to be parallel to $\om$ and, moreover, consider only $\mathbf k$ parallel to them to simplify our discussion. In this case the only preferred direction in space is along $\BB$, so $h_{\pm}$ could depend only on the absolute value of the component of momentum orthogonal to $\BB$ so the Lorentz force term in ~(\ref{reduced})  vanishes and we have
\begin{equation}
-i\nu h_{\pm} + i\dot{\xx}\cdot\kk h_{\pm} = -\frac{1}{\tau}(h_{\pm}\mp\bar h) .
\end{equation}
So,
\begin{eqnarray}
\langle h_+\rangle_+ = \left<\frac{\bar h}{-i\nu\tau + i\dot{\xx}\cdot\kk\tau +1}\right>_+ , \\
\langle h_-\rangle_- = \left<\frac{-\bar h}{-i\nu\tau + i\dot{\xx}\cdot\kk\tau +1}\right>_- .
\end{eqnarray}
The condition of consistency of this system is
\begin{multline}{\label{consist}}
\left<\frac{1}{-i\nu\tau + i\dot{\xx}\cdot\kk\tau +1}\right>_+ + \left<\frac{1}{-i\nu\tau + i\dot{\xx}\cdot\kk\tau +1}\right>_- \\ = -\chi .
\end{multline}
We now expand $\nu$ in series assuming that $B$ and $\omega$ are small:
\begin{equation}
\nu(\vec k) = \nu_{n}(k) + \nu_{a}(\vec k) + ... ,
\end{equation}
where $\nu_{n} = O(B^0,\omega^0)$ ("normal") is the term in the absence of any magnetic field and vorticity and $\nu_{a} = O(B^1,\omega^1)$ ("anomalous") is the first term of expansion which is responsible for the phenomenon we are studying. If $\BB = 0$, $\om = 0$:
\begin{multline}{\label{relax+}}
\left<\frac{\tau^{-1}}{-i\nu_{n} + i\dot{\xx}\cdot\kk + \tau^{-1}}\right>_+ \\
= \frac{\tau^{-1}}{2}\int_{-1}^1\frac{d\cos\theta}{-i\nu_{n} +ik\cos\theta+\tau^{-1}}\int_{\pp}\partial_p f_{0+} \\
 = \frac{1}{2ik\tau}\log\left(\frac{1+i(k-\nu_{n})\tau}{1-i(k+\nu_{n})\tau}\right)\int_{\pp}\sqrt{G_+}\partial_p f_{0+}  .
\end{multline}
Let us denote $\chi_+ = \int_{\pp}\sqrt{G_+}\partial_p f_{0+}$ and, analogously $\chi_- = \int_{\pp}\sqrt{G_-}\partial_p f_{0-}$. From (\ref{susc}) we have
\begin{equation}
\chi_+ +\chi_- = -\chi .
\end{equation}
From (\ref{consist}), (\ref{relax+}) and the analogous equation for antifermions we have
\begin{equation}
\nu_{n}=\frac{1-k\tau\cot(k\tau)}{i\tau}\approx-i\tau\left(\frac{1}{3}(k\tau)^2+...\right) ,
\end{equation}
so the first term of the expansion is diffusion. Let now $\BB , \om\neq 0$
\begin{multline}
\left<\frac{\tau^{-1}}{-i\nu + i\dot{\vec x}\cdot\vec k + \tau^{-1}}\right>_+ \\
 = \frac{1}{(2\pi)^2}\int_{-1}^1d\cos\theta
\int_0^{\infty}\frac{p^2\sqrt G_+\tau^{-1}\partial f_{0+}/\partial p dp}{-i\nu + \frac{ik}{\sqrt G_+}(\cos\theta +\frac{B_+'}{2p^2})+\tau^{-1}} \\
\approx \frac{\chi_+}{2ik\tau}\log\left(\frac{1+i(k-\nu)\tau}{1-i(k+\nu)\tau}\right)
+\frac{\tau^{-1}}{2(2\pi)^2}\int_0^{\infty}B_+'\frac{\partial f_{0+}}{\partial p} dp \\
\times\int_{-1}^1d\cos\theta\frac{\cos\theta(\tau^{-1}-i\nu)-ik+2(\tau^{-1}-i\nu)^2}{-i\nu + ik+\tau^{-1}} \\
=  \chi_+\left(1+\frac{\nu_{a}}{2k}\frac{2ik\tau}{1-2i\nu_{n}\tau +(k^2-\nu_{n}^2)\tau^2}\right) \\
+ \frac{\tau^{-1}}{2(2\pi)^2}(-\frac{B}{2}-\frac{2\omega}{\beta}\log(1+e^{\beta\mu_0})) \frac{6\nu_{n}\tau}{k} .
\end{multline}
The last computation requires some clarifications. After the second equality sign we kept only terms up to first order in $B$ and $\omega$. Then in the zero order term we kept both $\nu_{n}$ and $\nu_{a}$ while in the first order term we kept only $\nu_{n}$ as $\nu_{a}$ is first order in $B$ and $\omega$ itself. Analogously
\begin{multline}
\left<\frac{\tau^{-1}}{-i\nu + i\dot{\vec x}\cdot\vec k + \tau^{-1}}\right>_- \\
=\chi_-\left(1+\frac{\nu_{a}}{2k}\frac{2ik\tau}{1-2i\nu_{n}\tau +(k^2-\nu_{n}^2)\tau^2}\right) \\
+ \frac{\tau^{-1}}{2(2\pi)^2}(-\frac{B}{2}+\frac{2\omega}{\beta}\log(1+e^{-\beta\mu_0})) \frac{6\nu_{n}\tau}{k} .
\end{multline}
So we get
\begin{equation}
\frac{i\tau\nu_{a}\chi}{1-2i\tau\nu_{n}+(k^2-\nu_{n}^2)\tau^2} - \frac{6\nu_{n}(B+2\omega\mu_0)}{8\pi^2k} = 0 ,
\end{equation}
and inserting $\nu_{n}$ we get
\begin{equation}
\nu_{a} = k\frac{B + 2\mu_0\omega}{4\pi^2\chi}\frac{3(1-k\tau\cot(k\tau))}{\sin^2(k\tau)} = k\frac{B + 2\mu_0\omega}{4\pi^2\chi} ,
\end{equation}
with the last equality holding for $k\tau\ll1$. Note that this anomalous (first order in $B$ and $\omega$) part of the dispersion relation turns out to be purely real and, hence, non-dissipative.

\section{Discussion}

In this paper we have taken hydrodynamic and kinetic approaches to find the velocity of gapless excitation in chiral media in presence of magnetic field and vorticity - Chiral Magnetic-Vortical Wave. Using these approaches we have rederived the result of (\cite{ref:CHW}) that in the first order in $B$ and $\omega$ the velocity in question is given just by the vector sum of velocities of CMW and CVW(in RTA we have proven that only for the case of parallel $\BB$ and $\om$ and the wave vector $\kk$ parallel to them as well). Another interesting result is that in RTA CMVW turns out to be non-dissipative in linear order in $B$ and $\omega$ as dispersion relation for $\nu_{a}$ turns out to be purely real.

\section{Acknowledgements}

Author is grateful to A.~S.~Gorsky for suggesting this problem and valuable discussions and to A.~Aristova for useful communications.

\begin{thebibliography}{99}

\bibitem{ref:CME} 
  K.~Fukushima, D.~E.~Kharzeev and H.~J.~Warringa,
{\sl ``The Chiral Magnetic Effect''},
  Phys.\ Rev.\ D {\bf 78}, 074033 (2008) [arXiv:0808.3382].
\bibitem{ref:CSE1}
  D.~T.~Son and A.~R.~Zhitnitsky,
{\sl ``Quantum anomalies in dense matter''},
  Phys.\ Rev.\ D {\bf 70}, 074018 (2004)
  [hep-ph/0405216].
\bibitem{ref:CSE2}
  M.~A.~Metlitski and A.~R.~Zhitnitsky,
  {\sl ``Anomalous axion interactions and topological currents in dense matter''},
  Phys.\ Rev.\ D {\bf 72}, 045011 (2005) [hep-ph/0505072].
\bibitem{ref:CVE1} 
N.~Banerjee, J.~Bhattacharya, S.~Bhattacharyya, S.~Dutta, R.~Loganayagam and P.~Surowka,
{\sl ``Hydrodynamics from charged black branes''},
JHEP {\bf 1101}, 094 (2011)
[arXiv:0809.2596].
\bibitem{ref:CVE2} 
  J.~Erdmenger, M.~Haack, M.~Kaminski and A.~Yarom,
  {\sl ``Fluid dynamics of R-charged black holes''},
  JHEP {\bf 0901}, 055 (2009)
  [arXiv:0809.2488].
\bibitem{ref:CVE3} 
  D.~T.~Son and P.~Surowka,
  {\sl ``Hydrodynamics with Triangle Anomalies''},
  Phys.\ Rev.\ Lett.\  {\bf 103}, 191601 (2009) 
  [arXiv:0906.5044].
\bibitem{ref:CVE4} 
K.~Landsteiner, E.~Megias and F.~Pena-Benitez,
{\sl ``Gravitational Anomaly and Transport''},
Phys.\ Rev.\ Lett.\  {\bf 107}, 021601 (2011)  
[arXiv:1103.5006].
\bibitem{ref:CMW}
  D.~E.~Kharzeev and H.~U.~Yee,
  {\sl ``Chiral Magnetic Wave''},
  Phys.\ Rev.\ D {\bf 83}, 085007 (2011)
  [arXiv:1012.6026].
\bibitem{ref:CVW}
  Y.~Jiang, X.~G.~Huang and J.~Liao,
{\sl ``Chiral vortical wave and induced flavor charge transport in a rotating quark-gluon plasma''},
  arXiv:1504.03201 [hep-ph].
\bibitem{ref:CHW}
  M.~N.~Chernodub,
  {\sl ``Chiral Heat Wave and wave mixing in chiral media''},
  arXiv:1509.01245 [hep-th].
\bibitem{ref:Kinetic1} 
  M.~A.~Stephanov, H.~U.~Yee and Y.~Yin,
  {\sl ``Collective modes of chiral kinetic theory in a magnetic field''},
  Phys.\ Rev.\ D {\bf 91}, 125014 (2015) [arXiv:1501.00222].
\bibitem{ref:Kinetic2}  
  M.~A.~Stephanov, Y.~Yin,
  {\sl ``Chiral Kinetic Theory''}
  Phys. Rev. Lett. {\bf 109},162001 (2012)[arXiv:1207.0747]
\end
{thebibliography}
 \end{document}